\documentclass{aa}  

\usepackage{graphicx}
\usepackage{txfonts}
\usepackage{xcolor}
\usepackage{hyperref}
\hypersetup{colorlinks, linkcolor=blue, citecolor=blue, urlcolor=blue}
\usepackage{amsmath}
\usepackage{ulem}

\newcommand{\pivec}{\mbox{\boldmath $\pi$}}

\newcommand{\te}{t_{\rm E}}
\newcommand{\thetae}{\theta_{\rm E}}

\newcommand{\pie}{\pi_{\rm E}}

\newcommand{\dl}{D_{\rm L}}

\definecolor{brown}{rgb}{0.59, 0.29, 0.0}
\definecolor{darkgreen}{rgb}{0.0, 0.42, 0.24}
\definecolor{darkblue}{rgb}{0.01, 0.31, 0.59}
\definecolor{darkblue}{rgb}{0.0, 0.25, 0.42}
\definecolor{blue}{rgb}{0.0,0.0,1.0}
\definecolor{green}{rgb}{0.0,1.0,0.0}

\begin{document}

\title{Three microlensing planets with no caustic-crossing features}

\author{
     Cheongho~Han\inst{1} 
\and Andrzej~Udalski\inst{2} 
\and Doeon~Kim\inst{1}
\and Chung-Uk~Lee\inst{3} 
\\
     Michael~D.~Albrow\inst{4}   
\and Sun-Ju~Chung\inst{3,5}      
\and Andrew~Gould\inst{6,7}      
\and Kyu-Ha~Hwang\inst{3} 
\and Youn~Kil~Jung\inst{3} 
\and Yoon-Hyun~Ryu\inst{3} 
\and In-Gu~Shin\inst{3} 
\and Yossi~Shvartzvald\inst{8}    
\and Jennifer~C.~Yee\inst{9}      
\and Weicheng~Zang\inst{10}       
\and Sang-Mok~Cha\inst{3,11} 
\and Dong-Jin~Kim\inst{3} 
\and Hyoun-Woo~Kim\inst{3} 
\and Seung-Lee~Kim\inst{3,5} 
\and Dong-Joo~Lee\inst{3} 
\and Yongseok~Lee\inst{3,11} 
\and Byeong-Gon~Park\inst{3,5} 
\and Richard~W.~Pogge\inst{7}
\\
(The KMTNet Collaboration),\\
     Przemek~Mr{\'o}z\inst{2,12} 
\and Micha{\l}~K.~Szyma{\'n}ski\inst{2}
\and Jan~Skowron\inst{2}
\and Rados{\l}aw~Poleski\inst{2} 
\and Igor~Soszy{\'n}ski\inst{2}
\and Pawe{\l}~Pietrukowicz\inst{2}
\and Szymon~Koz{\l}owski\inst{2} 
\and Krzysztof~Ulaczyk\inst{13}
\and Krzysztof~A.~Rybicki\inst{2}
\and Patryk~Iwanek\inst{2}
\and Marcin~Wrona\inst{2}
\and Mariusz Gromadzki\inst{2}
\\
(The OGLE Collaboration)\\
}

\institute{
     Department of Physics, Chungbuk National University, Cheongju 28644, Republic of Korea  \\ \email{\color{blue} cheongho@astroph.chungbuk.ac.kr}     
\and Astronomical Observatory, University of Warsaw, Al.~Ujazdowskie 4, 00-478 Warszawa, Poland                                                          
\and Korea Astronomy and Space Science Institute, Daejon 34055, Republic of Korea                                                                        
\and University of Canterbury, Department of Physics and Astronomy, Private Bag 4800, Christchurch 8020, New Zealand                                     
\and Korea University of Science and Technology, 217 Gajeong-ro, Yuseong-gu, Daejeon, 34113, Republic of Korea                                           
\and Max Planck Institute for Astronomy, K\"onigstuhl 17, D-69117 Heidelberg, Germany                                                                    
\and Department of Astronomy, The Ohio State University, 140 W. 18th Ave., Columbus, OH 43210, USA                                                       
\and Department of Particle Physics and Astrophysics, Weizmann Institute of Science, Rehovot 76100, Israel                                               
\and Center for Astrophysics $|$ Harvard \& Smithsonian 60 Garden St., Cambridge, MA 02138, USA                                                          
\and Department of Astronomy, Tsinghua University, Beijing 100084, China                                                                                 
\and School of Space Research, Kyung Hee University, Yongin, Kyeonggi 17104, Republic of Korea                                                           
\and Division of Physics, Mathematics, and Astronomy, California Institute of Technology, Pasadena, CA 91125, USA                                        
\and Department of Physics, University of Warwick, Gibbet Hill Road, Coventry, CV4 7AL, UK                                                               
}
\date{Received ; accepted}

\abstract
{}
{
We search for microlensing planets with signals exhibiting no caustic-crossing features, considering 
the possibility that such signals may be missed due to their weak and featureless nature. 
}
{
For this purpose, we reexamine the lensing events found by the KMTNet survey before the 2019 season. 
From this investigation, we find two new planetary lensing events, KMT-2018-BLG-1976 and KMT-2018-BLG-1996.  
We also present the analysis of the planetary event OGLE-2019-BLG-0954, for which the planetary signal was 
known, but no detailed analysis has been presented before.  We identify the genuineness of the planetary 
signals by checking various interpretations that can generate short-term anomalies in lensing light curves.  
}
{
From Bayesian analyses conducted with the constraint from available observables, we find that the 
host and planet masses are 
$(M_1, M_2)\sim (0.65~M_\odot, 2~M_{\rm J})$ for KMT-2018-BLG-1976L, 
$\sim (0.69~M_\odot,  1~M_{\rm J})$ for KMT-2018-BLG-1996L, and 
$\sim (0.80~M_\odot, 14~M_{\rm J})$ for OGLE-2019-BLG-0954L.
The estimated distance to OGLE-2019-BLG-0954L, $3.63^{+1.22}_{-1.64}$~kpc, indicates that it is located 
in the disk, and the brightness expected from the mass and distance  matches well the brightness of the 
blend, indicating that the lens accounts for most of the blended flux.  The lens of OGLE-2019-BLG-0954 
could be resolved from the source by conducting high-resolution follow-up observations in and after 2024.
}
{}


\maketitle

\begin{table*}[htb]
\small
\caption{Coordinates and fields\label{table:one}}
\begin{tabular}{lccccc}
\hline\hline
\multicolumn{1}{c}{Event}                      &
\multicolumn{1}{c}{(RA, decl.)$_{\rm J2000}$}  &
\multicolumn{1}{c}{$(l, b)$}                   &
\multicolumn{1}{c}{Field}                      \\
\hline
KMT-2018-BLG-1976      & (17:45:25.12, -35:42:01.19)    &  $(-5^\circ\hskip-2pt .802, -3^\circ\hskip-2pt .483)$     &  KMT37                     \\
KMT-2018-BLG-1996      & (17:54:42.84, -22:49:13.48)    &  $(6^\circ\hskip-2pt .299, 1^\circ\hskip-2pt .382)  $     &  KMT38                     \\
OGLE-2019-BLG-0954     & (17:51:39.18, -29:36:39.60)    &  $(0^\circ\hskip-2pt .100, -1^\circ\hskip-2pt .475) $     &  OGLE501, KMT02, KMT42     \\
\hline
\end{tabular}
\end{table*}

\section{Introduction}\label{sec:one}

The microlensing signal of a planet is characterized by a short-term anomaly appearing on the smooth
single-lens single-source (1L1S) light curve produced by the host of the planet \citep{Mao1991, Gould1992b}.  
The signal is produced because the planet induces caustics, denoting the source positions at which the 
lensing magnification of a point source becomes infinite. The majority of microlensing planets reported 
so far\footnote{The Extrasolar Planets Encyclopaedia (http://exoplanet.eu/)} have been detected 
through caustic-crossing features, which are produced by passage of a source over the planet-induced caustic.  
However, planetary signals can still be generated without a caustic crossing of the source.  This is 
because the magnification excess induced by a planet extends to a considerable region beyond the caustic, 
and thus a planetary signal can be produced by the approach, instead of crossing, of the source to the 
caustic: ``non-caustic-crossing channel'', for example, OGLE-2016-BLG-1067Lb \citep{Calchi2019}.  The 
non-caustic-crossing channel is important because its cross-section to a planetary signal is substantially 
larger than the caustic size.  \citet{Zhu2014} estimated that in a KMT-like survey, about half of all 
detectable events should lack caustic features.

Despite the large cross section, planets detected through the non-caustic-crossing channel comprise a
minor fraction of all reported microlensing planets. The relative rarity of planets discovered through 
the non-caustic-crossing channel is mostly attributed to the difficulty of detecting planetary signals. 
Two factors contribute to this difficulty.  First, planetary signals with no caustic-crossing features 
are likely to be weak.  Due to the nature of the caustic, the planetary signal produced by the caustic 
crossing is usually strong, although the strength varies depending on the source size. In contrast, the 
strength of the non-caustic-crossing signal is much weaker because the source flux does not go through 
a great magnification induced by the caustic.  For the same reason, planetary signals produced by caustic 
crossings can be missed if the caustic-crossing features are not covered.  Second, planetary signals 
produced by the non-caustic-crossing channel tend to be featureless. Caustic-crossing planetary signals 
usually exhibit characteristic features, such as the caustic-crossing spikes and the U-shape trough 
between the spikes, and this helps one to easily notice the signal. On the contrary, non-caustic-crossing 
signals, in most cases, do not exhibit a noticeable feature that specifies the planetary origin of the 
signal.

Another reason for the lack of planet reports detected through the non-caustic-crossing channel is rooted 
in the difficulty of finding scientific issues that might draw attention.  In most cases, important 
scientific issues for discovered planets are drawn from their physical parameters, such as the mass and 
distance. For microlensing planets, these parameters can be determined by measuring extra observables in 
addition to the basic observable of the lensing event timescale $\te$. These extra observables are the 
microlens parallax, $\pie$, and the angular Einstein radius, $\thetae$. The microlens parallax is 
measurable for long timescale events, in which the lensing light curves exhibit deviations from a symmetric 
form due to the orbital motion of Earth around the Sun \citep{Gould1992}. The angular Einstein radius is 
measurable for caustic-crossing planetary events, in which the light curve during the caustic crossing 
exhibits deviations from a point-source form due to finite-source effects. For planets detected through 
the non-caustic-crossing channel, however, it is difficult to measure $\thetae$ because the lensing light 
curve is not subject to finite-source effects.  This makes it difficult to uniquely measure the physical 
parameters of the detected planetary system. For this reason, some planetary lensing events are left 
without detailed analyses even after planetary signals are noticed.  Nevertheless, it is important to 
report all detected planets for the construction of a complete planet sample, from which the planet 
frequency and demographic properties are deduced.

In this paper, we report the third result from the project that has been conducted by reinvestigating the
data collected by the Korea Microlensing Telescope Network \citep[KMTNet:][]{Kim2016} survey 
before the 2019 season with the aim of finding unrecognized planetary signals. In the first part of the
project, \citet{Han2020b} reexamined lensing events associated with faint source stars, considering the
possibility that planetary signals in these events might be missed due to the large photometric uncertainty. 
From this work, they reported four unnoticed or unpublished microlensing planets, including KMT-2016-BLG-2364Lb, 
KMT-2016-BLG-2397Lb, OGLE-2017-BLG-0604Lb, and OGLE-2017-BLG-1375Lb.  In the second part of the project, 
\citet{Han2021} reported a super-Earth planet orbiting a very low-mass star, KMT-2018-BLG-1025Lb, found from 
the systematic inspection of high-magnification microlensing events in the previous data. The result that we 
report in this paper comes from the third part of the project. In this work, we reinvestigate lensing events 
to search for microlensing planets with no caustic-crossing features.  From this investigation, we find two 
planetary lensing events, KMT-2018-BLG-1976 and KMT-2018-BLG-1996, for which the planetary signals were not 
noticed before. We also present the analysis of the planetary event OGLE-2019-BLG-0954, for which the planetary 
signal was known, but no detailed analysis has been presented before.

For the presentation of the analysis, we organize the paper as follows.  In Sect.~\ref{sec:two}, we describe 
the observations of the analyzed events and the data acquired from the observations. In Sect.~\ref{sec:three}, 
we describe details of the modeling conducted to explain the observed anomalies in the lensing light curves.  
In Sect.~\ref{sec:four}, we characterize the source stars of the events, and we test the possibility of 
constrain $\thetae$.  In Sect.~\ref{sec:five}, we estimate the physical lens parameters using the available 
observables.  We discuss some of the implications of these results in Sect.~\ref{sec:six} and conclude in 
Sect. ~\ref{sec:seven}.

\section{Observation and data}\label{sec:two}

The three planetary lensing events KMT-2018-BLG-1976, KMT-2018-BLG-1996, and OGLE-2019-BLG-0954
commonly occurred on stars located toward the Galactic bulge field. The equatorial and galactic
coordinates of the individual events are listed in Table~\ref{table:one}.

The events KMT-2018-BLG-1976 and KMT-2018-BLG-1996 were found from the post-season searches for
lensing events in the data collected during the 2018 season by the KMTNet survey using the Event
Finder System algorithm \citep{Kim2018}. While KMT-2018-BLG-1976 and KMT-2018-BLG-1996 were
found solely by the KMTNet survey, the event OGLE-2019-BLG-0954/KMT-2019-BLG-3289 was found by
two surveys, first by the Optical Gravitational Lensing Experiment \citep[OGLE:][]{Udalski2015} 
survey and later by the KMTNet survey. Hereafter, we designate this event as OGLE-2019-BLG-0954 
according to the chronological order of the discoveries.  The observations by the KMTNet survey 
were conducted using the three identical telescopes that are globally located in three continents: the 
Siding Spring Observatory in Australia (KMTA), the Cerro Tololo Interamerican Observatory in Chile 
(KMTC), and the South African Astronomical Observatory in South Africa (KMTS). The aperture of each 
KMTNet telescope is 1.6 meter, and the field of view of the camera mounted on the telescope is 
4~deg$^2$. The observations by the OGLE survey were done using the 1.3~m telescope located at the 
Las Campanas Observatory in Chile. The OGLE telescope is equipped with a camera yielding 1.4~deg$^2$ 
field of view.

For both surveys, the images of the source stars were obtained mainly  in the $I$ band, and a fraction 
of images were acquired in the $V$ band for the source color measurement. We will describe the detailed
procedure of the source color measurements in Sect.~\ref{sec:four}. The observational cadence varies 
depending on the events and surveys. The events KMT-2018-BLG-1976 and KMT-2018-BLG-1996 were located 
in the KMT37 and KMT38 fields, respectively, and both fields were observed with a 2.5~hr cadence. The 
event OGLE-2019-BLG-0954 was located in the OGLE501 and KMT02+KMT42 fields, which were observed 
with the cadences of 1 hour and 15 minute, respectively.  We note that KMT-2018-BLG-1976 is not in the 
OGLE footprint and that KMT-2018-BLG-1996 lies in the OGLE field BLG642, which was not observed for 
microlensing purposes after 2016.

Reduction of data and photometry of the events were carried out using the software pipelines developed
by the individual survey groups: \citet{Albrow2009} for KMTNet and \citet{Wozniak2000} for OGLE. Both 
of these photometry codes are based on the difference imaging technique \citep{Tomaney1996, Alard1998}, 
which is optimized for dense-field photometry. Following the routine described in \citet{Yee2012}, we 
readjust error bars of the data estimated from the automatized pipelines first to account for the scatter 
of the data, and second to make $\chi^2$ per degree of freedom for each data set become unity.

\section{Analyses}\label{sec:three}

The light curves of the three events analyzed in this work commonly exhibit weak anomalies with short
durations. In order to reveal the origins of the anomalies, we test three models under the 1L1S, 2L1S,
1L2S interpretations. The 2L1S modeling is done under the interpretation that the lens is a binary
object, while the 1L2S modeling is conducted under the interpretation that the source is a binary. The
2L1S modeling is done because a short-term anomaly can be produced by a planetary companion to the
lens.  The 1L2S model is tested because a subset of  1L2S events can produce short anomalies that 
mimic planetary signals \citep{Gaudi1998}.

A lensing light curve is described by different sets of parameters depending on the interpretation. 
A 1L1S lensing light curve is described by three parameters $(t_0, u_0, \te)$, which denote the 
peak time at the closest lens-source approach, the lens-source separation at $t_0$ (impact parameter), 
and the event time scale, respectively. The event time scale is defined as the time required for the 
source to cross the angular Einstein radius of the lens, that is, $\te= \thetae/\mu$, where $\mu$ denotes 
the relative lens-source proper motion. Modeling a 2L1S light curve requires additional parameters to 
describe the binarity of the lens.  These extra parameters are $(s, q, \alpha)$, which indicate the 
projected binary lens separation (scaled to $\thetae$) and mass ratio between the lens components, 
$M_1$ and $M_2$, and the angle between the source trajectory and $M_1$--$M_2$ axis (source trajectory 
angle), respectively.  For the description of the lensing light curve with an anomaly caused by a caustic 
crossing, during which the light curve is affected by finite-source effects, it is required to include an 
extra parameter $\rho$, which is defined as the ratio of the angular source radius $\theta_*$ to the angular 
Einstein radius, that is $\rho=\theta_*/\thetae$ (normalized source radius).  Describing the source binarity 
also requires one to include additional parameters $(t_{0,2}, u_{0,2}, \rho_2, q_F)$, where the first three 
are the closest time and separation between the second source and the lens, normalized source radius of the 
second source, and the last parameter denotes the flux ratio between the source stars. In the following 
subsections, we present details of the analyses conducted for the individual events.

\begin{table*}[htb]
\small
\caption{Lensing parameters of KMT-2018-BLG-1976\label{table:two}}
\begin{tabular}{lccccc}
\hline\hline
\multicolumn{1}{c}{Parameter}     &
\multicolumn{1}{c}{2L1S (Close)}  &
\multicolumn{1}{c}{2L1S (Wide)}   &
\multicolumn{1}{c}{1L2S}          &
\multicolumn{1}{c}{1L1S}          \\
\hline
$\chi^2$                    &  $2126.3             $   &  $2125.9            $    &  $2150.9            $    &  $2168.0            $     \\
$t_0$ (HJD$^\prime$)        &  $8183.083 \pm 0.048 $   &  $8183.112 \pm 0.049$    &  $8182.627 \pm 0.292$    &  $8183.087 \pm 0.043$     \\
$u_0$                       &  $0.14106 \pm 0.00477$   &  $0.146 \pm 0.005   $    &  $0.121 \pm 0.013   $    &  $0.142 \pm 0.005   $     \\
$\te$ (days)                &  $42.49 \pm 1.00     $   &  $41.76 \pm 1.01    $    &  $45.13 \pm 1.82    $    &  $42.34 \pm 1.04    $     \\
$s$                         &  $0.708 \pm 0.030    $   &  $1.227 \pm 0.064   $    &  --                      &  --                       \\
$q$ ($10^{-3}$)             &  $2.89 \pm 0.79      $   &  $3.13 \pm 0.95     $    &  --                      &  --                       \\
$\alpha$ (rad)              &  $1.065 \pm 0.027    $   &  $1.074 \pm 0.029   $    &  --                      &  --                       \\
$\rho$                      &  --                      &  --                      &  --                      &  --                       \\
$t_{0,2}$ (HJD$^\prime$)    &  --                      &  --                      &  $8190.922 \pm 2.415$    &  --                       \\
$u_{0,2}$                   &  --                      &  --                      &  $0.104 \pm 0.056   $    &  --                       \\
$\rho_2$                    &  --                      &  --                      &  --                      &  --                       \\
$q_F$                       &  --                      &  --                      &  $0.066 \pm 0.209   $    &  --                       \\
\hline
\end{tabular}
\end{table*}

\begin{figure}[t]
\includegraphics[width=\columnwidth]{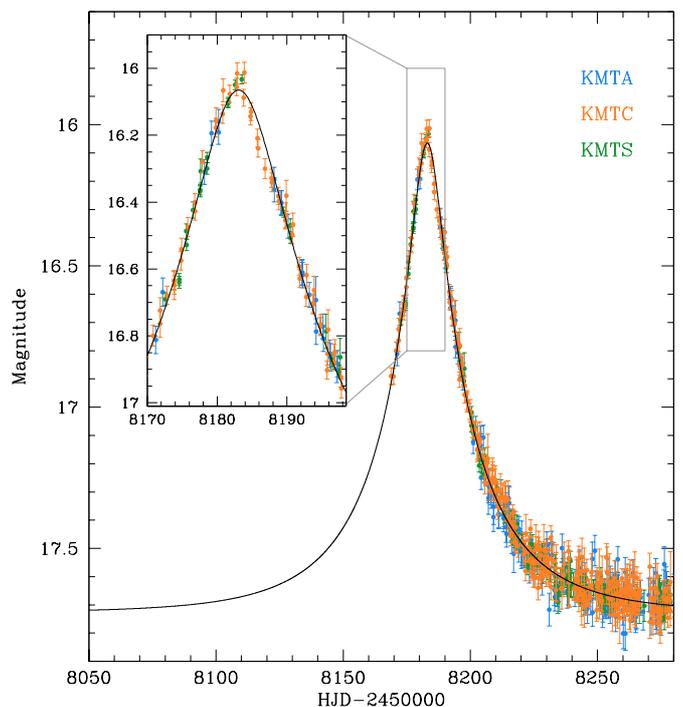}
\caption{
Lensing light curve of KMT-2018-BLG-1976. The inset shows the zoomed-in view of the peak region.
}
\label{fig:one}
\end{figure}

\subsection{KMT-2018-BLG-1976}\label{sec:three-one}

\begin{figure}
\includegraphics[width=\columnwidth]{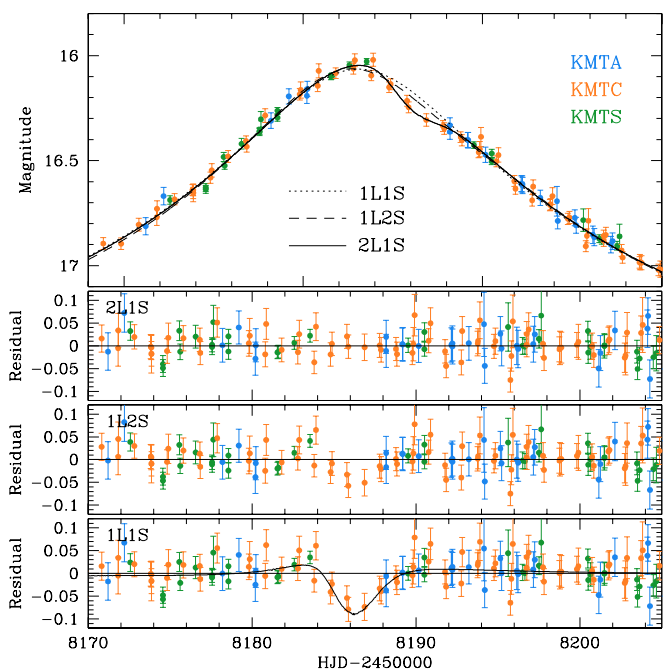}
\caption{
Enlarged view in the peak region of the KMT-2018-BLG-1976 light curve. The three lower panels show the 
residuals from the three tested models under 1L1S, 1L2S, and 2L1S interpretations. The 2L1S model is 
based on the wide binary-lens interpretation. The curve drawn in the 1L1S residual panel is the 
difference between the 2L1S and 1L1S models.
}
\label{fig:two}
\end{figure}

Figure~\ref{fig:one} shows the lensing light curve of the event KMT-2018-BLG-1976. It shows that the 
lensing-induced magnification of the source flux started during the time gap between the end of the 2017 
season and the beginning of the 2018 season.  The light curve reached its peak on 2018-03-05,
 ${\rm HJD}^\prime\equiv {\rm HJD}-2450000\sim8183$, at which the source became brighter by 
$\Delta I\sim 1.9$ magnitude than the baseline magnitude of $I_{\rm base}=18.02$, according to the KMTNet 
scale, and then gradually returned to its baseline.  At first glance, the light curve appears to be that of 
a 1L1S event. The 1L1S modeling yields the lensing parameters of $(u_0, \te)\sim (0.14, 42.3~{\rm days})$, 
indicating that the source flux was magnified by $A_{\rm peak}=(u_0+2)/[u_0(u_0^2+4)^{1/2}]\sim 7.2$ at 
the peak. The 1L1S model curve is drawn over the data points in Figure~\ref{fig:one}, and the full 
lensing parameters and their uncertainties are listed in Table~\ref{table:two}. However, a close inspection 
reveals that the light curve exhibits an anomaly that appears around the peak. In Figure~\ref{fig:two}, we 
present the enlarged view of the peak region along with the residuals from the 1L1S model. The anomaly, 
which lasted for about 6 days during the period of $8184\lesssim {\rm HJD}^\prime \lesssim 8190$, shows 
negative deviations in most of the anomaly region, but it exhibits a slight positive deviation at the 
beginning of the anomaly.  Although the deviation of the anomaly is small, $\lesssim 0.1$~mag, we consider 
the anomaly is significant because the data obtained by the different telescopes delineate a consistent 
pattern of deviation.

In order to explain the anomaly, we first test a 1L2S model.  The 1L2S modeling is done with the initial 
parameters of $(t_0, u_0, \te)$ obtained from the 1L1S modeling, and the initial values of the parameters 
related to the second source, $(t_{0,2}, u_{0,2}. \rho_2, q_F)$,
are assigned considering the magnitude and the location of the anomaly in the 
lensing light curve.  This modeling yields the lensing parameters of $(u_0, u_{0,2}, q_F)\sim (0.12, 0.10, 0.07)$, 
indicating that the second source, $S_2$, with a flux about 7\% of the flux from the primary source, $S_1$, 
approaches closer to the lens than $S_1$ does.  The full lensing parameters and their uncertainties of the 
1L2S model are listed in Table~\ref{table:two}, and the model curve and the residuals are shown in 
Figure~\ref{fig:two}.  The introduction of an extra source improves the fit by $\Delta\chi^2\sim 17.1$ 
with respect to the 1L1S model, reducing the depth of the negative deviations.  However, the model still 
leaves subtle but noticeable deviations, indicating that a different interpretation is needed.

\begin{figure}
\includegraphics[width=\columnwidth]{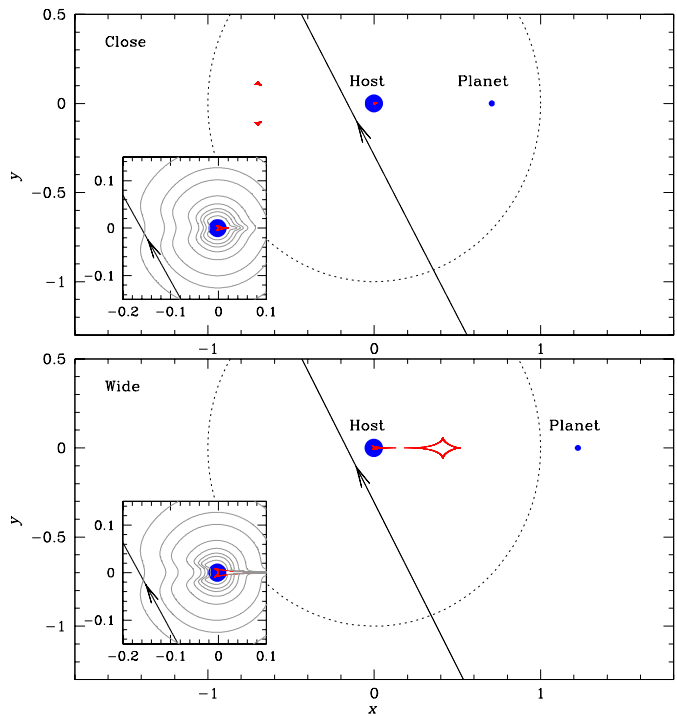}
\caption{
Lens system configuration of KMT-2018-BLG-1976. The line with an arrow indicates the source trajectory, 
the two blue dots marked by ``host'' and ``planet'' are lens positions, and red figures represent caustics.  
The dotted circle around the host represents the Einstein ring. The inset shows the zoomed-in view of the 
central magnification region, in which the grey curves around the position of the host represent equi-magnification 
contours The upper and lower panels are for the ``inner'' (close, $s < 1.0$) and ``outer'' (wide, $s > 1.0$) 
solutions, respectively.
}
\label{fig:three}
\end{figure}

We then test a 2L1S model. The 2L1S modeling is conducted in two steps. In the first step, we conduct 
a grid search for the binary lens parameters $(s, q)$, while the other parameters are searched for based on 
the downhill $\chi^2$ minimization approach using the Markov Chain Monte Carlo (MCMC) algorithm. This step 
enables us to find local solutions that are subject to various types of degeneracy. In the second step, 
we inspect the local solutions in the $s$--$q$ parameter space, and then refine them by allowing all 
parameters (including $s$ and $q$) to vary.  We list the best-fit lensing parameters of the 2L1S model 
in Table~\ref{table:two}.

We identify two 2L1S solutions, one with $s>1$ and the other with $s<1$. Very often, when two such
solutions are found, they are identified as, respectively, the ``wide'' and ``close'' degenerate solutions
that \citet{Griest1998} and  \citet{Dominik1999} identified for central caustics. In the present
case, however, this is not correct. This is actually the ``outer/inner'' degeneracy for planetary
caustics that was identified by \citet{Gaudi1997}.  From Figure~\ref{fig:three}, one can see that,
in both cases, the dip is produced by the source crossing the long trough that extends along the
planet-host axis on the opposite side of the planet. In the ``close'' (``inner'') solution, there is a
separate set of planetary caustics, and the source passes ``inside'' these, i.e., between these caustics
and the central caustic. In the ``wide'' (``outer'') solution, the planetary and central caustics have
merged into a single, so-called ```resonant'' caustic. However, it is still the case that there is a
long trough extending from the ``back end'' of the resonant caustic that retains substantial
structure from the previously-separate planetary caustics. The source passes ``outside'' the tips of
these structures.

\begin{figure}
\includegraphics[width=\columnwidth]{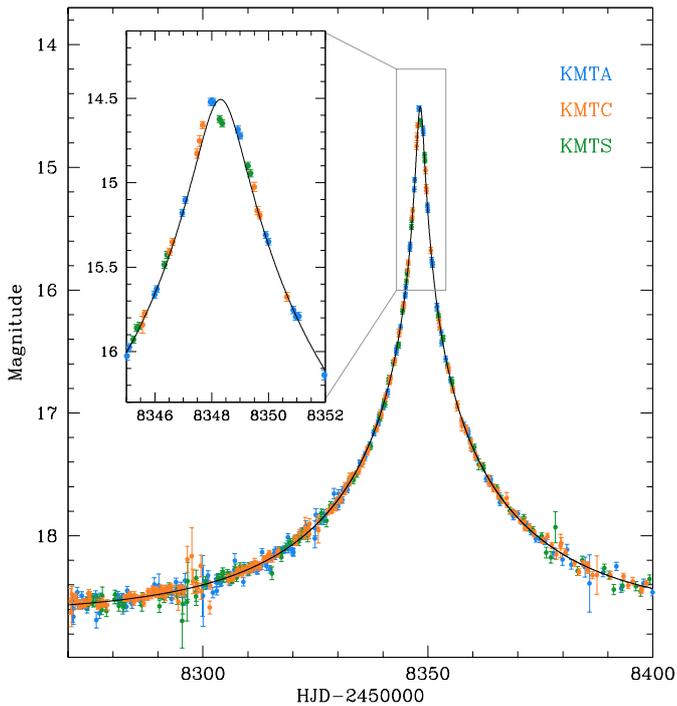}
\caption{
Light curve of KMT-2018-BLG-1996. The zoomed-in view of the peak region is shown in the inset.
}
\label{fig:four}
\end{figure}

\citet{Herrera2020} were the first to recognize that this pair of caustic morphologies is the ``inner/outer'' 
degeneracy. Indeed, the light curve of OGLE-2018-BLG-0677, which they analyzed, looks remarkably similar to 
that of KMT-2018-BLG-1976, except that the duration of the dip is much shorter. Closely related to this fact, 
the difference in the two values of $s$ is much smaller $\Delta s = (s_{\rm w} - s_{\rm c})/2 = 0.03$ (versus 
0.26 in the present case), and both values of $s$ were less than 1.0.  Here $s_{\rm w}$ and $s_{\rm c}$ denote 
the binary separations of the wide and close solutions, respectively.  The opposite to the present case was 
seen in OGLE-2016-BLG-1195 \citep{Bond2017, Shvartzvald2017}, in which a non-caustic ``bump'' was observed 
rather than a ``dip'', which yielded an inner/outer degeneracy but offset from the major-image planetary 
caustic (rather than minor-image caustic as for KMT-2018-BLG-1976). In that case also, $s_{\rm c}<1$ and 
$s_{\rm w}>1$. However, $\Delta s = 0.05$, a factor of 5 smaller than in the present case.

\citet{Yee2021} have conjectured that there is a continuous transition between the ``inner/outer''
planetary-caustic degeneracy of \citet{Gaudi1997} and the ``close/wide'' central-caustic degeneracy of 
\citet{Griest1998} and \citet{Dominik1999}. If correct, KMT-2018-BLG-1976 represents an extreme case along 
this continuum, with the largest $\Delta s$ of any event to date in the ``inner/outer'' regime.

The outer (wide) solution, with $s > 1.0$, is only marginally favored ($\Delta\chi^2=0.4$) over the inner 
(close) solution, with $s < 1.0$.  The model curve and the residual of the 2L1S model (for the wide solution) 
are shown in Figure~\ref{fig:two}.  It is found that the 2L1S model well describes the anomaly, in both 
negative- and positive-deviation regions, improving the fit by $\Delta\chi^2=42.1$ and 17.1 with respect 
to the 1L1S and 1L2S models, respectively.  For both solutions, the estimated binary mass ratio is 
$q\sim 3\times 10^{-3}$, indicating that the companion to the primary lens is a planetary-mass object.  
Considering the fairly long time scale, $\te\sim 42.5$~days, of the event, we check the feasibility of 
measuring $\pie$. From the modeling considering microlens-parallax effects, we find that it is difficult 
to securely determine $\pie$ due to the incomplete coverage of the rising-side light curve combined with 
the relatively large photometric errors.

\begin{table*}[htb]
\small
\caption{Lensing parameters of KMT-2018-BLG-1996\label{table:three}}
\begin{tabular}{lccccc}
\hline\hline
\multicolumn{1}{c}{Parameter}     &
\multicolumn{1}{c}{2L1S (Close)}  &
\multicolumn{1}{c}{2L1S (Wide)}   &
\multicolumn{1}{c}{1L2S}          &
\multicolumn{1}{c}{1L1S}          \\
\hline
$\chi^2$                    &  $1000.1            $    &  $979.6             $    &  $1034.2            $    &  $1183.4            $     \\
$t_0$ (HJD$^\prime$)        &  $8348.305 \pm 0.008$    &  $8348.305 \pm 0.008$    &  $8347.920 \pm 0.067$    &  $8348.316 \pm 0.007$     \\
$u_0$                       &  $0.018 \pm 0.001   $    &  $0.018 \pm 0.001   $    &  $0.019 \pm 0.002   $    &  $0.019 \pm 0.001   $     \\
$\te$ (days)                &  $47.08 \pm 0.56    $    &  $47.07 \pm 0.55    $    &  $47.89 \pm 0.72    $    &  $45.71 \pm 0.50    $     \\
$s$                         &  $0.672 \pm 0.041   $    &  $1.455 \pm 0.091   $    &  --                      &  --                       \\
$q$ ($10^{-3}$)             &  $1.69 \pm 0.38     $    &  $1.51 \pm 0.39     $    &  --                      &  --                       \\
$\alpha$ (rad)              &  $1.429 \pm 0.046   $    &  $1.452 \pm 0.046   $    &  --                      &  --                       \\
$\rho$                      &  --                      &  --                      &  --                      &  --                       \\
$t_{0,2}$ (HJD$^\prime$)    &  --                      &  --                      &  $8348.824 \pm 0.147$    &  --                       \\
$u_{0,2}$                   &  --                      &  --                      &  $-0.021 \pm 0.002  $    &  --                       \\
$\rho_2$                    &  --                      &  --                      &  --                      &  --                       \\
$q_F$                       &  --                      &  --                      &  $0.84 \pm 0.95     $    &  --                       \\
\hline
\end{tabular}
\end{table*}

\begin{figure}
\includegraphics[width=\columnwidth]{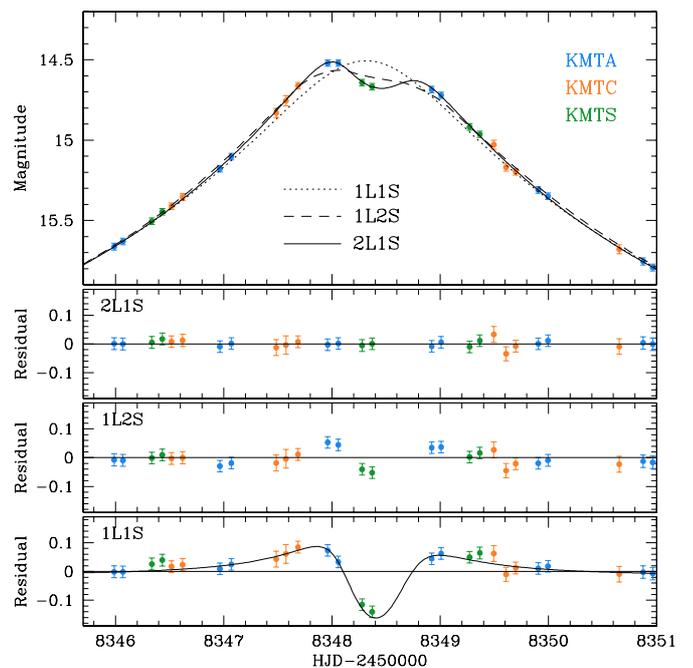}
\caption{
Zoomed-in view in the peak region of the KMT-2018-BLG-1996 light curve. Notations are same as those in 
Fig.~\ref{fig:two}.
The 2L1S model is for the wide solution with $s>1.0$.
}
\label{fig:five}
\end{figure}

\subsection{KMT-2018-BLG-1996}\label{sec:three-two}

The lensing light curve of KMT-2018-BLG-1996 is shown in Figure~\ref{fig:four}, in which the enlarged 
view around the peak region is shown in the inset.  The event shares a common characteristics with the 
KMT-2018-BLG-1976 in the sense that the light curve appears to be approximated by a 1L1S curve, and a 
short-lasting smooth anomaly appears around the peak.  The estimated lensing parameters from a 1L1S 
modeling are $(u_0, \te)\sim (0.019, 45.7~{\rm days})$, and thus the source flux was magnified by 
$A_{\rm peak}\sim 53$ at the peak. The full 1L1S lensing parameters are listed in Table~\ref{table:three}. 
The data near the peak exhibit deviations, which lasted for about 4 days, from the 1L1S model. We present the 
enlarged view of the peak region and the residuals from the 1L1S model in Figure~\ref{fig:five}.  The residuals 
of the 1L1S model exhibit a similar pattern to that of the event KMT-2018-BLG-1976: a central dip with negative 
deviations and bumps with positive deviations before and after the central dip.

In order to find the origin of the anomaly, we test both 1L2S and 2L1S models. The lensing parameters 
of the solutions from these modelings are listed in Table~\ref{table:three} along with the $\chi^2$ values 
of the model fits.  As in the case of KMT-2018-BLG-1976, the 2L1S modeling yields two solutions, and the 
degeneracy between the two solutions is very severe with $\Delta\chi^2=0.08$.  The 2L1S solutions provide 
better fits than the 1L1S and 1L2S models with $\Delta\chi^2=203.8$ and 54.6, respectively.  The estimated 
binary lens parameters are $(s, q)\sim (0.67, 1.69\times 10^{-3})$ for the close solution and 
$\sim (1.46, 1.51\times 10^{-3})$ for the wide solution, and thus the companion to the lens is a planetary 
mass object regardless of the solution.  In Figure~\ref{fig:five}, we present the model curves and residuals 
from the 1L2S and 2L1S (wide) models.

\begin{figure}
\includegraphics[width=\columnwidth]{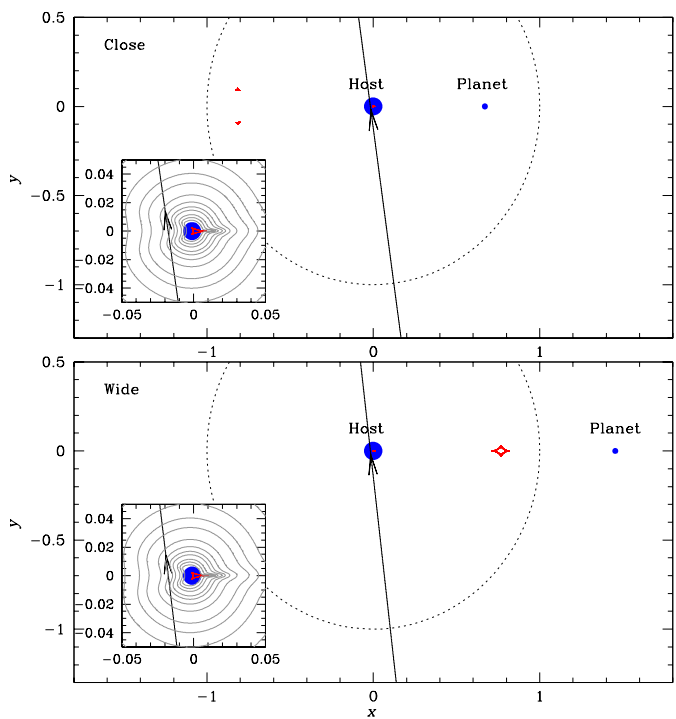}
\caption{
Lens system configuration of
KMT-2018-BLG-1996. Notations are same
as those in Fig.~\ref{fig:three}.
}
\label{fig:six}
\end{figure}

\begin{figure}
\includegraphics[width=\columnwidth]{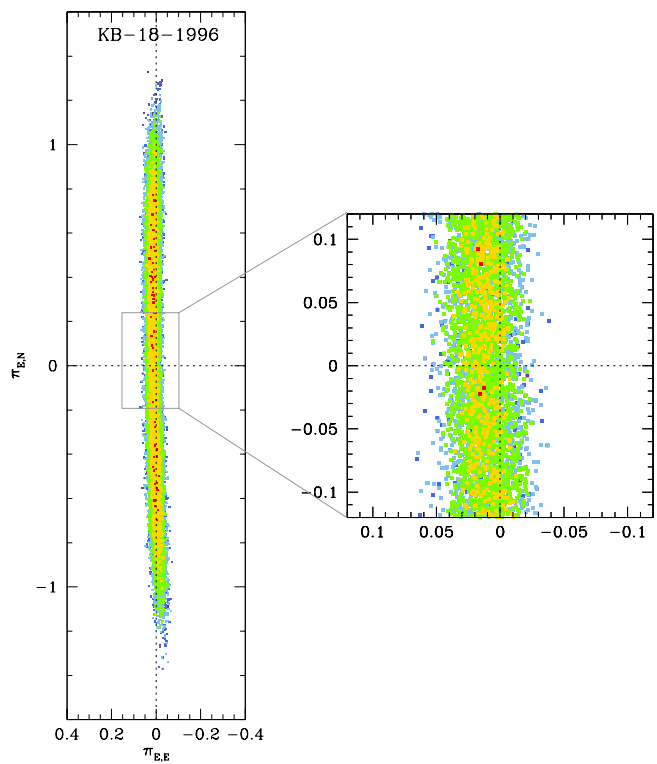}
\caption{
Scatter plot of points in the MCMC chain on the $\pi_{{\rm N},E}$--$\pi_{{\rm N},N}$ parameter plane obtained 
from the 2L1S modeling of the KMT-2018-BLG-1996 light curve considering higher-order effects.  Red, yellow, 
green, cyan, and blue colors are used to denote points with $\leq1\sigma$, $\leq 2\sigma$, $\leq3\sigma$, 
$\leq4\sigma$, and $\leq5\sigma$, respectively.  
}
\label{fig:seven}
\end{figure}

In the upper and lower panels of Figure~\ref{fig:six}, we present the lens system configurations of the 
close and wide 2L1S solutions, respectively.  From the comparison of the configurations with those of 
KMT-2018-BLG-1976 presented in Figure~\ref{fig:three}, it is found that the origin of the anomaly is 
very similar to that of KMT-2018-BLG-1976 in the sense that it arises from the source crossing the 
magnification trough that extends out from the host in the direction opposite to the planet.  The 
configuration of KMT-2018-BLG-1996 shows two important differences from that of KMT-2018-BLG-1976. 
First, the lens-source impact parameter, $u_0\sim 0.018$, is substantially smaller than that of KMT-2018-BLG-1976, 
with  $u_0\sim 0.14$.  Thus, in contrast to OGLE-2018-BLG-1976, the anomaly is restricted to the immediate 
vicinity of the central caustic, where we expect the tight mathematical correspondence between the tidal 
expansion (of the wide solution) and quadrupole expansion (of the close solution) to hold 
\citep{Dominik1999, An2005}.  Indeed, in contrast to OGLE-2018-BLG-1976, the central caustics are nearly 
identical for the two solutions and $s_w\times s_c = 0.98$, that is, quite close to unity.  Second, the 
source passes the binary axis with a steeper source trajectory angle, $\alpha\sim 82^\circ$, and this 
results in well developed positive deviations at both sides of the negative deviation region, while only 
a single bump is seen in the anomaly of KMT-2018-BLG-1976, for which $\alpha\sim 61^\circ$.

In light of the relatively long time scale, $\te\sim 47$~days, of the event and the relatively good photometry 
of the observed light curve, we check the possibility of measuring $\pie$ by conducting an additional modeling 
considering the microlens-parallax effect.  In the modeling, we also take account of the lens-orbital effect 
that may correlate with the microlens-parallax effect \citep{Batista2011, Skowron2011}.  From this modeling, 
we find that it is difficult to constrain $\pie$ for two reasons.  First, the fit improvement with the 
consideration of the higher-order effects, $\Delta\chi^2 <1.0$, is negligible.  Second, the uncertainty of 
the measured $\pie$ is too big to constrain the physical lens parameters.  This is shown in Figure~\ref{fig:seven}, 
in which we present the scatter plot of points in the MCMC chain  on the $\pi_{{\rm E},E}$--$\pi_{{\rm E},N}$ 
parameter plane.  Here $\pi_{{\rm E},E}$ and $\pi_{{\rm E},N}$ denote the east and north components of the 
microlens-parallax vector $\pivec_{\rm E}$.    In the plot, the red, yellow, green, cyan, and blue colors 
are used to denote points with $\leq1\sigma$, $\leq 2\sigma$, $\leq3\sigma$, $\leq4\sigma$, and $\leq5\sigma$, 
respectively.  The scatter plot shows that the model with higher-order effects is consistent with a static 
model, and the uncertainty of $\pie$ measurement, especially the north component of the parallax vector is 
too big for the meaningful constraint of the physical lens parameters.

\begin{table*}[htb]
\small
\caption{Lensing parameters of OGLE-2019-BLG-0954\label{table:four}}
\begin{tabular}{lccccc}
\hline\hline
\multicolumn{1}{c}{Parameter}     &
\multicolumn{1}{c}{2L1S}          &
\multicolumn{1}{c}{1L2S}          &
\multicolumn{1}{c}{1L1S}          \\
\hline
$\chi^2$                    &    $9260.4            $    &  $9684.3            $    &  $10245.4           $     \\
$t_0$ (HJD$^\prime$)        &    $8662.096 \pm 0.140$    &  $8661.991 \pm 0.104$    &  $8662.121 \pm 0.107$     \\
$u_0$                       &    $0.426 \pm 0.014   $    &  $0.429 \pm 0.005   $    &  $0.403 \pm 0.004   $     \\
$\te$ (days)                &    $28.54 \pm 0.68    $    &  $25.98 \pm 0.33    $    &  $28.00 \pm 0.11    $     \\
$s$                         &    $0.737 \pm 0.006   $    &  --                      &  --                       \\
$q$                         &    $0.017 \pm 0.002   $    &  --                      &  --                       \\
$\alpha$ (rad)              &    $0.373 \pm 0.028   $    &  --                      &  --                       \\
$\rho$                      &    $\leq 0.0013       $    &  --                      &  --                       \\
$t_{0,2}$ (HJD$^\prime$)    &    --                      &  $8674.368 \pm 0.062$    &  --                       \\
$u_{0,2}$                   &    --                      &  $-0.001 \pm 0.003  $    &  --                       \\
$\rho_2$                    &    --                      &  --                      &  --                       \\
$q_F$                       &    --                      &  $0.005 \pm 0.001$       &  --                       \\
\hline
\end{tabular}
\end{table*}

\begin{figure}
\includegraphics[width=\columnwidth]{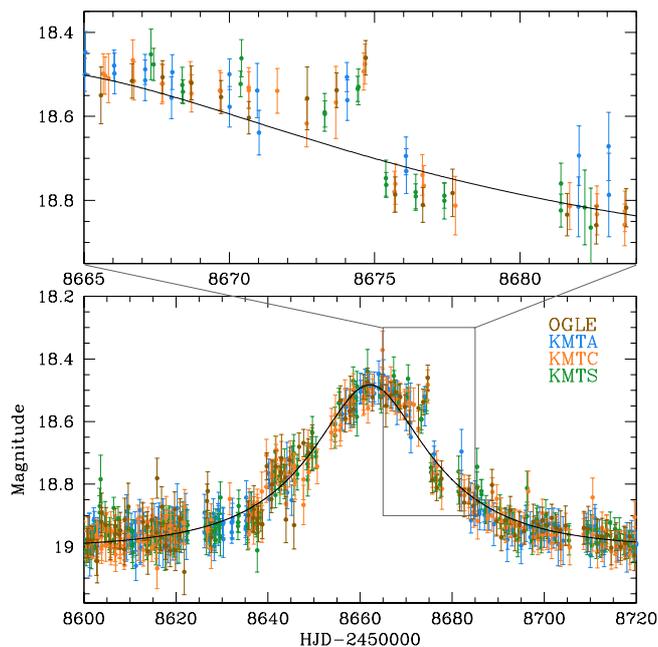}
\caption{
Light curve of OGLE-2019-BLG-0954.  The inset shows the enlargement of the anomaly region.
}
\label{fig:eight}
\end{figure}

\begin{figure}
\includegraphics[width=\columnwidth]{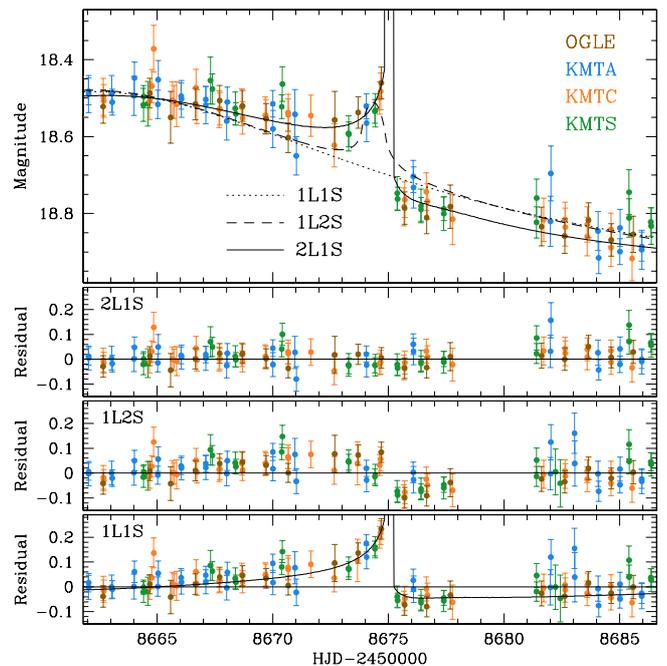}
\caption{
Model curves and residuals of the 1L1S, 1L2S, and 2L1S solutions in the region around the anomaly of 
OGLE-2019-BLG-0954. Notations are same as those in Fig.~\ref{fig:two},
except that the range of the residuals, -0.15 -- 0.29 in magnitudes, is asymmetric to better show the residual 
from the 1L1S model.
}
\label{fig:nine}
\end{figure}

\subsection{OGLE-2019-BLG-0954}\label{sec:three-three}

Figure~\ref{fig:eight} shows the lensing light curve of the event OGLE-2019-BLG-0954. 
We note that the photometric quality of the data is low due to the faintness of the source, 
and thus we binned the data with a 12 hr interval to better show the anomalous nature of the light 
curve.  The upper panel shows the zoomed-in view of the region around the anomaly centered at the 
bump at ${\rm HJD}^\prime\sim 8675$. The anomaly exhibits $\sim 0.3$~mag deviation from a 1L1S model, 
which is marked by a solid curve drawn over the data points. The anomaly lasted for about 7 days,
and it displays both positive (before the bump) and negative (after the bump) deviations. The light 
curve is similar to those of the previous two events in the sense that the anomaly does not exhibit 
a prominent caustic-crossing feature.  The major difference is that the anomaly appears not around 
the peak region but on the falling side of the light curve about $\sim 13$~days after the peak at 
$t_0\sim 8662$. The magnification of the event is low with $A\sim 2.4$, and the apparent source 
brightness at the peak is about 0.5~mag brighter than the baseline with $I_{\rm base}\sim 19.01$.  
The impact parameter of the source trajectory estimated from a 1L1S modeling is $u_0\sim 0.4$. In 
Table~\ref{table:four}, we list the 1L1S lensing parameters.

In order to explain the anomaly, we test both the 1L2S and 2L1S models.  We note that the modeling 
is done with all data, not using the binned data, although the light curve in Figure~\ref{fig:eight} 
is shown with binned data.  The full lensing parameters of these solutions are listed in 
Table~\ref{table:four}, and the model curves and residuals  around the anomaly region are shown in 
Figure~\ref{fig:nine}.  Like the previous two events, it is found that the 2L1S model with a 
planetary-mass companion provides a better fit than the 1L1S and 1L2S models, with $\Delta\chi^2=423.9$ 
and 985.0, respectively. The estimated planet/primary mass ratio between the lens components is 
$q\sim 0.017$, indicating that the mass of the lens companion is in the planetary regime.  The event 
is different from the two previous events in two aspects. First, the anomaly was produced by the 
caustic crossing of a source, although the caustic-crossing feature was not covered by the data.  
Second, the solution is uniquely determined without any degeneracy.  As we will mention below, the 
anomaly was produced by the source star's crossing over the planetary caustic induced by a planet 
with $s<1.0$.  In this case, there is, in general, no degeneracy between the solutions with $s<1.0$ 
and $s>1.0$, because the planetary caustics induced by a close and a wide planet are different from 
each other both in the number and shape \citep{Han2006}.

\begin{figure}
\includegraphics[width=\columnwidth]{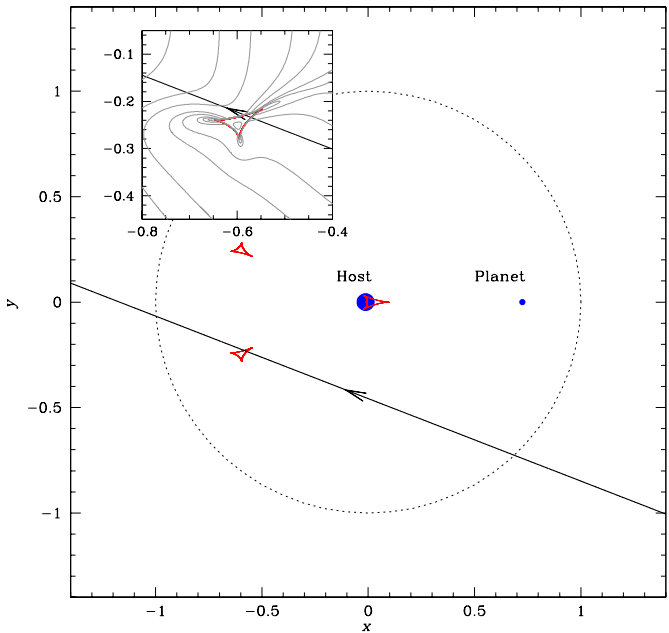}
\caption{
Lens system configuration of OGLE-2019-BLG-0954. Notations are same as those in Fig.~\ref{fig:three}.
}
\label{fig:ten}
\end{figure}

\begin{figure}
\includegraphics[width=\columnwidth]{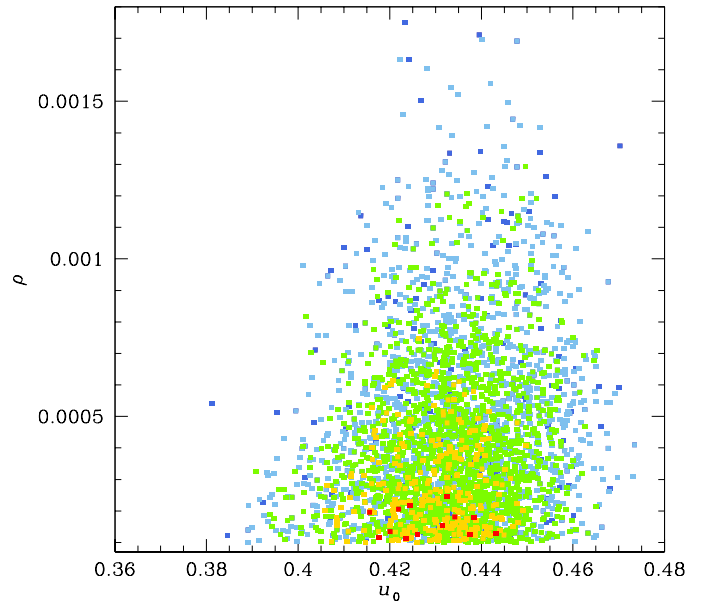}
\caption{
Scatter plot of points in the MCMC chain on the $u_0$--$\rho$ parameter plane obtained from the 
2L1S modeling of the OGLE-2019-BLG-0954 lensing light curve.  The colors of the points are defined 
in the same way as in Fig.~\ref{fig:seven}.
}
\label{fig:eleven}
\end{figure}

Figure~\ref{fig:ten} shows the configuration of the lens system. It shows that the planetary signal 
was produced by the source crossing over one of the two sets of the tiny planetary caustics produced by 
the planet with a normalized separation $s\sim 0.74$.  The duration of the caustic crossings, as measured 
by the time gap between the caustic entrance and exit, is $\sim 7.7$~days according to the model. This 
time corresponded to the night in Australia, at which the sky was clouded out and thus no observation was 
conducted.  Although not resolved, the caustic crossings of the source are supported by the positive 
deviation, that lasted about 4.5 days during $8670.0 \lesssim {\rm HJD}^\prime \lesssim 8674.5$, before 
the caustic crossings, and the negative deviation, that lasted $\sim 2.5$~days during $8675.5 \lesssim 
{\rm HJD}^\prime \lesssim 8678.0$, after the caustic crossings.  See the curve of the difference between 
the 2L1S and 1L1S models presented in the bottom panel of Figure~\ref{fig:nine}, which shows the rapid 
rise and fall of the 1L1S residual around the times of the caustic crossings.

Another important difference of OGLE-2019-BLG-0954 from the other events is that it is possible to place 
an upper limit on the normalized source radius despite the fact that the detailed caustic-crossing feature 
was not resolved by the data.  This can be seen in Figure~\ref{fig:eleven}, where we present the scatter 
plot of MCMC points on the $u_0$--$\rho$ plane.  It shows that the upper limit of the normalized source 
radius is $\rho_{\rm max}\sim 1.3\times 10^{-3}$ as measured at 3$\sigma$.  As we will show in 
Sect.~\ref{sec:five}, the upper limit of $\rho$ provides an important constraint on the location of the lens.

\begin{figure}
\includegraphics[width=\columnwidth]{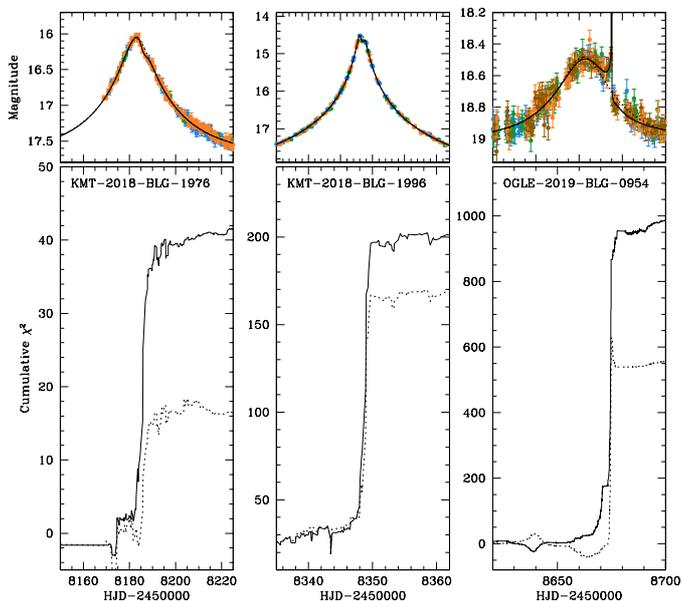}
\caption{
Cumulative distributions of $\Delta\chi^2$  with respect to the 1L1S models for the individual lensing 
events. The solid and dotted distributions are
$\Delta\chi_{\rm 2L1S}^2 = \chi_{\rm 1L1S}^2 - \chi_{\rm 2L1S}^2$
and
$\Delta\chi_{\rm 1L2S}^2 = \chi_{\rm 1L1S}^2 - \chi_{\rm 1L2S}^2$,
respectively. 
}
\label{fig:twelve}
\end{figure}

In Figure~\ref{fig:twelve}, we present the cumulative distributions of $\Delta\chi^2$ with respect 
to the 1L1S models for the individual lensing events. The light curves in the upper panels are inserted 
to show the region of the fit improvement. For each event, we present two cumulative distributions, where 
the solid and dotted curves represent 
$\Delta\chi_{\rm 2L1S}^2 = \chi_{\rm 1L1S}^2 - \chi_{\rm 2L1S}^2$ 
and $\Delta\chi_{\rm 1L2S}^2 = \chi_{\rm 1L1S}^2 - \chi_{\rm 1L2S}^2$, 
respectively. Although the strength 
of the planetary signal, $40 \lesssim \Delta\chi^2 \lesssim 985$, varies depending on the events, the 
distributions show that the improvement of the fit occurs at the time of the anomaly, indicating that 
the anomaly is a short-term perturbation, which can be explained either by a 2L1S and a 1L2S model. The 
fact that the 2L1S solutions provide better fits than the 1L2S solutions clearly indicates that the 
anomalies are of planetary origin.

\begin{table*}[htb]
\small
\caption{Source properties of three events\label{table:five}}
\begin{tabular}{llllcc}
\hline\hline
\multicolumn{1}{c}{Quantity}     &
\multicolumn{1}{c}{KMT-2018-BLG-1976}          &
\multicolumn{1}{c}{KMT-2018-BLG-1996}          &
\multicolumn{1}{c}{OGLE-2019-BLG-0954}         \\
\hline
$(V-I, I)$                &    $(2.106 \pm 0.036, 18.242 \pm 0.004)$    &  $(2.54 \pm 0.07, 18.82 \pm 0.02)   $    &   $(3.36 \pm 0.12, 20.32 \pm 0.07)    $   \\
$(V-I, I)_{\rm RGC}$      &    $(2.298, 15.972)                    $    &  $(3.665, 17.137)                   $    &   $(3.665, 17.137)                    $   \\
$(V-I, I)_0$              &    $(0.868 \pm 0.036, 16.991 \pm 0.004)$    &  $(0.994 \pm 0.071, 15.952 \pm 0.016$    &   $(0.755 \pm 0.115, 17.613 \pm 0.068)$   \\
$\theta_*$ ($\mu$as)      &    $1.50 \pm 0.12                      $    &  $2.80 \pm 0.280                    $    &   $0.994 \pm 0.133                    $   \\
$\thetae$ (mas)           &    --                                       &  --                                      &   $\geq 0.8$                              \\
$\mu$ (mas yr$^{-1}$)     &    --                                       &  --                                      &   $\geq 10.2$                             \\
\hline
\end{tabular}
\end{table*}

\section{Source stars}\label{sec:four}

In this section, we specify the source stars of the events. In general, the main purpose of the source 
characterization is to estimate the angular Einstein radius, which helps to better constrain the physical 
lens parameters. For the measurement of $\thetae$, it is required to measure the normalized source radius 
$\rho$, from which the angular Einstein radius is measured by $\thetae=\theta_*/\rho$, with the angular 
source radius estimated from the source color and brightness. Then, the prerequisite for the $\thetae$ 
estimation is to measure $\rho$. For KMT-2018-BLG-1976 and KMT-2018-BLG-1996, the $\rho$ parameters cannot 
be determined because the lensing light curves do not exhibit finite-source deformations due to the absence 
of caustic-crossing features. For OGLE-2019-BLG-0954, however, it is possible to place the lower limit on 
$\thetae$, because the upper limit of $\rho$ is constrained, that is, $\theta_{\rm E,min}=\theta_*/\rho_{\rm max}$. 
Although $\thetae$ values cannot be constrained for the events KMT-2018-BLG-1976 and KMT-2018-BLG-1996, we 
specify their source stars for the sake of completeness.

\begin{figure}
\includegraphics[width=\columnwidth]{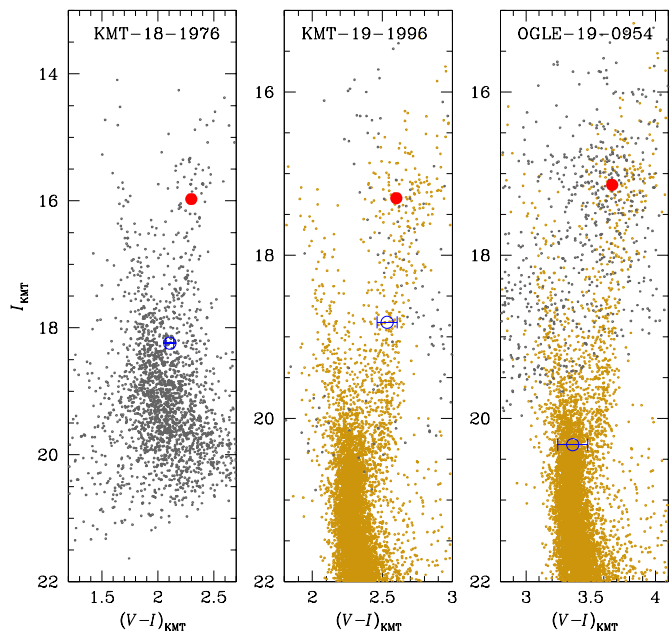}
\caption{
Source positions (blue empty dots) in
the instrumental color-magnitude diagrams
(CMDs). The red filled dot indicate the centroid
of the red giant clump. For KMT-2018-BLG-1996
and OGLE-2019-BLG-0954, CMDs from the {\it HST}
observations (brown dots) are additionally
presented.
}
\label{fig:thirteen}
\end{figure}

Figure~\ref{fig:thirteen} shows the locations of the source stars (blue empty circles with error bars) 
of the individual events in the instrumental color-magnitude diagrams (CMDs) of neighboring stars around 
the source stars constructed using the pyDIA photometry \citep{Albrow2017} of the KMTC data sets. Also 
marked are the centroids of the red giant clump (RGC, red filled dots). The instrumental source color, 
$V-I$, and brightness, $I$, are measured from the regression of $V$- and $I$-band data with the variation 
of the lensing magnification. For KMT-2018-BLG-1996 and OGLE-2019-BLG-0954, the quality of the $V$-band 
data are not good enough to reliably determine the source colors, although the $I$-band brightness is securely 
determined.  In order to estimate the source colors for these events, we apply the method of \citet{Bennett2008} 
using the {\it Hubble Space Telescope} ({\it HST}) CMD.  In this method, the two sets of CMDs constructed from 
the ground-based and {\it HST} observations \citep{Holtzman1998} are aligned using the RGC centroid, and then 
the source color is estimated as that of a star on either the main-sequence or the giant branch of the {\it HST} 
CMD considering the $I$-band brightness difference between the source and RGC centroid. With the measured 
instrumental source color and brightness, the reddening and extinction corrected values, $(V-I, I)_0$, are 
estimated from the offsets in color and brightness between the source and RGC centroid, $\Delta(V-I, I)$, 
and using the RGC centroid as a reference \citep{Yoo2004} by 
\begin{equation} 
(V-I, I)_0 = (V-I, I)_{\rm RGC,0} + \Delta (V-I, I),  
\label{eq1}
\end{equation} 
where $(V-I, I)_{\rm RGC,0}$ denotes the de-reddened color and brightness of the RGC centroid, which are known 
from \citet{Bensby2013} and \citet{Nataf2013}, respectively.  In Table~\ref{table:five}, we list the 
estimated the values $(V-I, I)$, $(V-I, I)_{\rm RGC}$, and $(V-I, I)_0$ for the individual events. For 
OGLE-2019-BLG-0954, we additionally present the lower limits of the angular Einstein radius and the relative 
lens-source proper motion, $\mu$. The value of $\mu$ is estimated from the combination of $\thetae$ and $\te$ 
by $\mu =\thetae/\te$.

\section{Physical lens parameters}\label{sec:five}

Although the microlens parallax is measurable for none of the analyzed events, it is still possible to
constrain the physical parameters of the lens mass and distance based on the event timescale, because 
it is related to the physical lens parameters by
\begin{equation}
\te={\thetae \over \mu};\ \ \ 
\thetae = (\kappa M_{\rm tot}\pi_{\rm rel})^{1/2};\ \ \ 
\pi_{\rm rel} = {\rm AU}\left( {1\over D_{\rm L}} - {1\over D_{\rm S}}\right).
\label{eq2}
\end{equation}
Here 
$\kappa=4G/(c^2{\rm AU})$, $M_{\rm tot}=M_1+M_2$, and 
$D_{\rm L}$ and $D_{\rm S}$ denote the distances to the lens and source, respectively. For OGLE-2019-BLG-0954,
the measured lower limit of $\thetae$ can place an additionally constraint on the lens parameters.

\begin{figure}
\includegraphics[width=\columnwidth]{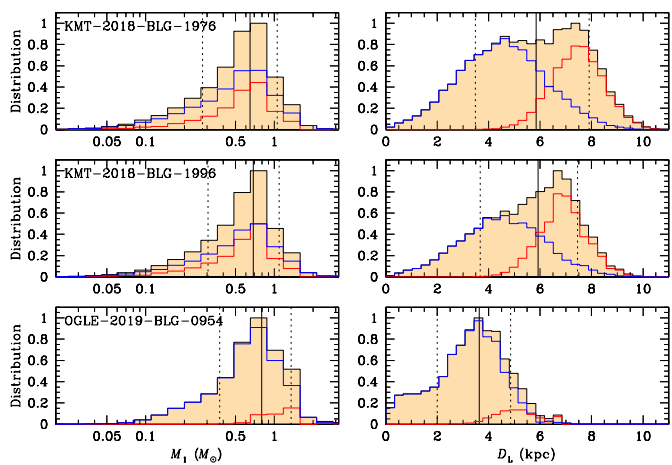}
\caption{
Bayesian posteriors of the planet host mass (left panels) and distance (right panels) for the three events: 
KMT-2018-BLG-1976L, KMT-2018-BLG-1996L, and OGLE-2019-BLG-0954L. In each panel, the blue and red curves 
represent the contributions by the disk and bulge lens populations, respectively, and the black curve is 
the sum of individual lens populations. The vertical solid line indicates the median of the distribution, 
and the two dotted lines represent the 1$\sigma$ range of the distribution.
}
\label{fig:fourteen}
\end{figure}

We estimate the physical lens parameters by conducting a Bayesian analysis using the available constraints 
for the individual events. In the Bayesian analyses, we produce a large number ($2\times 10^7$) of lensing 
events by conducting a Monte Carlo simulation using a prior Galactic model. The Galactic model is defined 
by the mass function, physical and dynamical distributions of Galactic objects.  For the mass function of 
lens objects, we adopt the model defined in \citet{Jung2018}. For the physical distribution, we adopt the 
\citet{Robin2003} model for disk objects, and the \citet{Han2003} model for bulge objects.  For the dynamical 
distribution, we adopt the \citet{Jung2021} model, that is constructed based on the Gaia catalog 
\citep{Gaia2016, Gaia2018} for bulge objects, and the modified \citet{Han1995} model for disk objects. 
For more details about the Galactic model, see \citet{Jung2021}.

Figure~\ref{fig:fourteen} shows the posterior distributions for the mass of the primary lens, $M_1$, and 
the distance to the lens obtained from the Bayesian analysis.  In each panel, the contributions by the disk 
and bulge lens populations are marked by blue and red curves, respectively, and the black curve is the 
combined contribution by the two lens populations. We mark the median value and 1$\sigma$ range of the 
distribution by a vertical solid and two dotted lines, respectively. The uncertainty range of each parameter 
is estimated as the 16\% and 84\% of the distribution.

\begin{table*}[htb]
\small
\caption{Physical lens parameters of three events\label{table:six}}
\begin{tabular}{llllcc}
\hline\hline
\multicolumn{1}{c}{Lens}                   &
\multicolumn{1}{c}{}                   &
\multicolumn{1}{c}{$M_1$ ($M_\odot$)}      &
\multicolumn{1}{c}{$M_2$ ($M_{\rm J}$)}    &
\multicolumn{1}{c}{$D_{\rm L}$ (kpc)}      &
\multicolumn{1}{c}{$a_\perp$ (AU)}         \\
\hline
KMT-2018-BLG-1976L  &  (Close)  & $0.65^{+0.41}_{-0.37}$       &  $1.96^{+1.23}_{-1.12}$    &   $5.84^{+2.06}_{-2.36}$  &  $2.04^{+0.72}_{-0.83}$  \\
                    &  (wide)   & --                           &  $2.22^{+1.34}_{-1.21}$    &   --                      &  $3.54^{+1.25}_{-1.43}$  \\
\hline              
KMT-2018-BLG-1996L  &  (Close)  & $0.69^{+0.40}_{-0.38}$       &  $1.22^{+0.72}_{-0.63}$    &   $5.91^{+1.54}_{-2.24}$  &  $1.97^{+0.51}_{-0.25}$  \\
                    &  (wide)   & --                           &  $1.09^{+0.64}_{-0.60}$    &   --                      &  $4.27^{+1.11}_{-1.62}$  \\
\hline
OGLE-2019-BLG-0954L &           & $0.80^{+0.55}_{-0.42}$       &  $14.2^{+9.9}_{-7.5}$      &   $3.63^{+1.22}_{-1.64}$  &  $2.65^{+0.89}_{-1.20}$  \\
\hline
\end{tabular}
\end{table*}

In Table~\ref{table:six}, we list the estimated masses of the host and planet, distance, and 
projected planet-host separation ($a_\perp=s D_{\rm L}\thetae$) for the individual lenses. For 
KMT-2018-BLG-1976L and KMT-2018-BLG-1996L, we present paired sets of the parameters corresponding to 
the close and wide solutions. The estimated host and planet masses are 
$(M_1, M_2)\sim (0.65~M_\odot, 2~M_{\rm J})$ for KMT-2018-BLG-1976L, 
$\sim (0.69~M_\odot,  1~M_{\rm J})$ for KMT-2018-BLG-1996L, and 
$\sim (0.80~M_\odot, 14~M_{\rm J})$ for OGLE-2019-BLG-0954L.  The estimated mass of OGLE-2019-BLG-0954L 
is in a good agreement with the $\thetae$--$M$ relation presented in Figure~7 of \citet{Kim2021}.  
According to the estimated masses, the hosts of planets are commonly K-type main-sequence stars.  
The fact that the lenses of all three events have similar masses can be understood by the similarity 
of the event time scales, ranging $\te\sim 25$--29~ days, which are major constraints on the determined 
physical parameters.  We note that the mass distribution of detected lenses is different from the mass 
function of stars in the solar neighborhood, in which M dwarfs are about 7 to 8 times more common than 
K dwarfs, because the lensing probability is proportional to the size of $\thetae$, which is proportional 
to $M^{1/2}$, and thus the lensing chance is higher for a lens with a higher mass.  The planets 
KMT-2018-BLG-1976Lb and KMT-2018-BLG-1996Lb are giant planets with masses similar to and about twice 
of the Jupiter mass, respectively.  On the other hand, the mass of OGLE-2019-BLG-0954Lb is at the 
planet/brown dwarf boundary.\footnote{There exist six cases of binary microlenses with companion masses 
at around this boundary including
MOA-2010-BLG-073L \citep{Street2013},
OGLE-2013-BLG-0102L \citep{Jung2015},
MOA-2015-BLG-337L \citep{Miyazaki2018},
OGLE-2016-BLG-1190L \citep{Ryu2018},
OGLE-2017-BLG-1375L \citep{Han2020b}, and
KMT-2019-BLG-1339L \citep{Han2020a}.} 
For all lenses, the planets are located beyond the snow lines of the hosts, regardless of the close or 
wide solutions.

We note that the OGLE-2019-BLG-0954L is very likely to be in the disk, while the chances for the lens to 
be in the disk and bulge are roughly alike for the other two events.  The constraint on the location of 
OGLE-2019-BLG-0954L is mostly given by the relatively large angular Einstein radius, which is 
$\gtrsim 0.8~{\rm mas}$, combined with the high relative lens-source proper motion, which is 
$\gtrsim 10.2~{\rm mas}~{\rm yr}^{-1}$.  Considering the close distance to OGLE-2019-BLG-0954L, 
$D_{\rm L}\sim 3.6$~kpc, we check the possibility that a significant fraction of the blended flux comes 
from the lens. The expected brightness of the lens estimated from its mass, $\sim 0.8~M_\odot$, and 
distance, $\dl\sim 3.6$~kpc, is $I\sim 19.4$, assuming that the extinction to the lens is about half of 
the extinction toward the source star of $A_I\sim 2.6$. This matches well the estimated brightness of the 
blend of $I_{\rm b}\sim 19.5$.  Therefore, it is plausible that the lens comprises most of the blended flux.  
Considering that the relative lens-source proper motion is very high, this can be confirmed if follow-up 
observations using high-resolution instrument are conducted in the near future.  Assuming that the separation 
for the lens-source resolution is $\sim 50$~mas, as demonstrated in the case of OGLE-2005-BLG-169 from the 
Keck AO observations conducted by \citet{Batista2015}, the lens could be resolved from the source in 2024.

\section{Discussion}\label{sec:six}

Among the three planets, one was detected by passage over a minor-image (triangular) planetary
caustic (OGLE-2019-BLG-0954), one by passage near a central caustic (KMT-2018-BLG-1996), and
one by passage relatively far from a ``resonant/near-resonant'' caustic (KMT-2018-BLG-1976). For
random source trajectories going through the Einstein ring, the cross sections of the first two type
are small because central caustics and minor-image planetary caustics are small. Indeed, it was
the small size of the planetary caustic in OGLE-2019-BLG-0954 that led to it entering our sample:
small gaps in relatively dense and continuous coverage meant that there was no coverage over the
short cusp crossing. However, as \citet{Yee2021} have pointed out, resonant and near-resonant (defined
as having ``magnification-deviation ridges'' of at least 10\% extending from the central to
planetary caustic) have much larger cross sections, and they account for a large fraction of all
planetary detections. See their Figure~11, which shows that roughly half of resonant/near-resonant
planets are from near-resonant caustic structures. However, for these near-resonant caustics, a
substantial fraction (of order half) of their cross-section is comprised of the ``10\% ridge'' rather
the caustics. The light curves without caustic features that result from these trajectories are more
likely to be missed.

Finally, we note that the comparison of the caustic diagrams of KMT-2018-BLG-1976 and KMT-2018-BLG-1996 
lends credence to the conjecture of Yee+2021 that inner/outer degeneracy of planetary caustics 
\citep{Gaudi1997} and the close/wide degeneracy of central caustics \citep{Griest1998, Dominik1999} are 
limiting cases of a single continuum of degeneracies. In both cases, the light curve is characterized 
by a post-peak ``dip'' due to passage over the magnification trough along the planet-host axis on the 
opposite side from the planet. Yet the degeneracy takes very different forms. For KMT-2018-BLG-1976 (for 
which the source passes much farther from the central-or-resonant caustic), the product 
$s_{\rm w}\times s_{\rm c}=0.87$ is very far from unity. This is more characteristic of the inner/outer 
degeneracy of planetary caustics, for which one expects $(s_{\rm w}\times s_{\rm c})^{1/2}= s^\dagger$, 
where $s^\dagger = |-u_{\rm anom}\pm (u_{\rm anom}^2 +4)^{1/2}|/2$ is the predicted value of $s$ based 
on the time of the anomaly $t_{\rm anom}$: $u_{\rm anom}^2 = u_0^2 + [(t_0-t_{\rm anom})/\te]^2$. 
For KMT-2018-BLG-1976, $u_{\rm anom} = 0.161$ and so $(s^\dagger)^2 = 0.85$, compared to the inner/outer 
``prediction'' of 0.87. By contrast, for KMT-2018-BLG-1996, $(s_{\rm w}\times s_{\rm c})^{1/2}=0.99$, 
very close to the prediction of unity for the close/wide regime.

\section{Summary}\label{sec:seven}

We reported the discoveries of 
three planetary microlensing events KMT-2018-BLG-1976 and KMT-2018-BLG-1996, and OGLE-2019-BLG-0954.  
The planets in these events were found from the systematic reinvestigation of the microlensing events 
found by the KMTNet survey before the 2019 season, which was conducted to search for weak planetary 
signals with no obvious caustic-crossing features.  Among the events,  the planetary signals in 
KMT-2018-BLG-1976 and KMT-2018-BLG-1996 were not noticed before, and the signal in OGLE-2019-BLG-0954 
was known, but no detailed analysis had been presented before.  We tested various interpretations to 
explain the observed short-term anomalies in the lensing light curves, and this confirmed that the 
signals were of planetary origin.  From the  Bayesian analyses, it was estimated that the host and planet 
have masses $(M_1, M_2)\sim (0.65~M_\odot, 2~M_{\rm J})$ for KMT-2018-BLG-1976L, 
$\sim (0.69~M_\odot,  1~M_{\rm J})$ for KMT-2018-BLG-1996L, and $\sim (0.80~M_\odot, 14~M_{\rm J})$ 
for OGLE-2019-BLG-0954L.  It turned out that the lens of OGLE-2019-BLG-0954 was located in the disk, 
and the its flux accounted for most of the blended flux.  We predict that OGLE-2019-BLG-0954L would 
be resolved from the source by conducting high-resolution follow-up observations in and after 2024.

\begin{acknowledgements}
Work by C.H. was supported by the grants of National Research Foundation of Korea
(2019R1A2C2085965 and 2020R1A4A2002885).
Work by A.G. was supported by JPL grant 1500811.
This research has made use of the KMTNet system operated by the Korea
Astronomy and Space Science Institute (KASI) and the data were obtained at
three host sites of CTIO in Chile, SAAO in South Africa, and SSO in
Australia.
The OGLE project has received funding from the National Science Centre, Poland, grant
MAESTRO 2014/14/A/ST9/00121 to AU.
\end{acknowledgements}

\end{document}